\documentstyle{article}

\newcommand{\mathbf}{\bf}

\begin{document}

\begin{center}
{\huge\bf A Model Of The Integer Quantum Hall Effect }
\end{center}

\vspace{1cm}
\begin{center}
{\large\bf
F.GHABOUSSI}\\
\end{center}

\begin{center}
\begin{minipage}{8cm}
Department of Physics, University of Konstanz\\
P.O. Box 5560, D 78434 Konstanz, Germany\\
E-mail: ghabousi@kaluza.physik.uni-konstanz.de
\end{minipage}
\end{center}

\vspace{1cm}

\begin{center}
{\large{\bf Abstract}}
\end{center}

\begin{center}
\begin{minipage}{12cm}
We discuss a model for the integer quantum Hall effect which is  
based on a Schroedinger-Chern-Simons-action functional for a  
non-interacting system of electrons in an electromagnetic field on a  
mutiply connected manifold. In this model the integer values of the  
Hall conductivity arises in view of the quantization of the  
Chern-Simons-action functional for electromagnetic potential.
\end{minipage}
\end{center}

\newpage
We discuss here an approach based on a new model of the integer  
quantum Hall-effect (IQHE) \cite{inteig}, according to which a  
direct quantization of an action functional for Ohm-equations of the  
IQHE results in quantization of the Hall-conductivity. The model is  
based on the following fact that a quantization presupposes an  
action of the system under consideration. Therefore, one needs first  
an action which must result in equations of motion describing the  
IQHE.

On the other hand the only equations which describe the IQHE are  
the well known Ohm-equations given by

\begin{equation}
            j_m = \sigma_H \epsilon_{nm}E_n \;, \; \epsilon_{mn} =  
- \epsilon_{nm} = 1 \;;\; m,n = 1,2   \;\;,
\end{equation}
\label{Hequat}

where the Hall-conductivity $\sigma_H =  
{\displaystyle\frac{ne}{B}}$ with $B := B_3$ is quantized in the  
units of ${\displaystyle\frac{e^2}{h}}$ \cite{inteig}. Here $n$ is  
the global carrier concentration on the sample.

Accordingly, we deduce that the Ohm-equations which describe the  
IQHE should be considered as the equations of motion which must  
result from some action.

There are various evidences to consider the 2+1-dimensional field  
theory to describe the IQHE:

On the one hand the dimension of the global density of carriers $n$  
is given empirically to be $L^{-2}$ in QHE, by the relation  
$\sigma_H = {\displaystyle\frac{ne}{B}}$. Thus, also the local  
density of carriers which is given theoretically according to the  
quantum mechanics by $\psi^*\psi$ must have dimension $L^{-2}$ (see  
also below).

On the other hand, in view of the $L^{-1}$-dimension of the  
electromagnetic potential the only possible minimal coupling between  
$\psi^*\psi$ and the electromagnetic potential can be considered in  
2+1-dimensions (see the action (3) below). We shall discuss the  
relation between the local and the global densities according to the  
constraint of the microscopic model.

\bigskip
We show here that a Schroedinger-Chern-Simons-action in  
2+1-dimensions for electrons in an electromagnetic field is the  
correct one to result in the Ohm-equations of motion (1). Therefore,  
a quantization of electromagnetic potential according to the  
Chern-Simons-action should result in the quantization of $\sigma_H$  
as it is manifested by the experimental results \cite{inteig}. The  
use of the Chern-Simons-term is forced in this model by the real and  
theoretical 2-dimensional picture of the IQHE and by the following  
chracters of this action which seems to be suitable to describe the  
IQHE.

On the one hand, the IQHE is known to be a bulk- or topological  
effect independent of the sample geometry.

On the other hand, the Chern-Simons-action is a topologically  
invariant theory whose observables are topological invariants of the  
base manifold. Thus, it is natural to apply the Chern-Simons-theory  
to explain the IQHE. This stand point is in good agreement with the  
already known topological approaches to IQHE [1][1c].

\medskip
Furthermore, the Chern-Simons-action describes the dynamics of a  
gauge potential which is according to the constraint of the theory a  
pure gauge potential, i. e. one with vanishing field strength.  
Thus, fortunately its line integrals in mutiply connected regions,  
i. e. with $(g\geq 1)$ which are typical for the IQHE, i. e. in  
non-simply connected regions as like as the Corbino-disc $( g = 1)$,  
disordered systems or regions with holes of impurities $(g\ge 1)$,  
are given by the topological invariants of the region. An example of  
this effect is the well known Bohm-Aharonov effect [1c]. These  
invariants depends only on the topology of the region or on its  
global structure in a manner which is belived to be manifested also  
by the experimental results of the IQHE \cite{inteig} [1c].

The quantization of such systems assignes also topological  
invariants to the quantum states of system \cite{schwarz}.

On the other hand, it is known that if one considers currents  
involved in the IQHE as the boundary currents, then most of  
experimental data can be understood in a satisfactory manner  
\cite{vkzwei}. It is a favour of the Chern-Simons-ansatz in a  
manifold with a spatial boundary that the boundary currents are the  
only allowed ones according to the constraint of the theory.

\bigskip
The Schroedinger-Chern-Simons-action of our model of IQHE, i. e.  
for the {\it non-interacting} 2-dimensional system of carriers in an  
elctromagnetic field, in $2+1$-dimensions which is defined on $M =  
\Sigma\times\mathbf R$ with $(g\geq 1)$ is given by (c = 1):

\begin{equation}
S = \frac{1}{4\pi}\int dt \int_{\Sigma} \psi^* [ i\hbar\partial_t -  
\frac{1}{2\mu} (-i\hbar\partial_m - e A_m )^2 - eA_0 ]\psi + h.c. -  
\frac{\sigma_H}{4\pi}
\int_M\epsilon^{\alpha\beta\gamma}A_{\alpha}\partial_{\beta}A_{\gamma}  
\;,
\end{equation}
\label{action}

where $\mu$ is the mass of electron and $A_{\alpha} (x_m,t)$ is  
still the cassical electromagnetic potential in 2+1-dimensions which  
must be quantized. Furthermore, everywhere is $\partial_{X} =  
{\displaystyle\frac{\partial}{\partial X}}$ and we consider (in  
accordance with the experimental arrangements of the QHE) that the  
$\Sigma$ has a boundary
 \cite{erklar}.

Moreover, we consider the disc $\Sigma$ as a non-simply connected  
region in accordance with the mentioned characteristics of samples  
used in the QHE-experiments.

Obviously, we use the $\sigma_H$ as the normalization parameter of  
the Chern-Simons-action to achive the Ohm-equations (1). It is  
justified to do so, because $\sigma_H$ can be considered as a  
dimensionless and locally constant quantity also in view of its well  
known topological or global character [1] \cite{diff}. Moreover, we  
suppressed the spin term within the usual Schroedinger-action for  
"spinors" in a magnetic field \cite{one} in view of the well known  
fact that the spin degenerecy is not essential for IQHE  
\cite{inteig} .

\bigskip
In view of the gauge freedom in the action, one should demand some  
gauge fixing condition to retain the true number of degrees of  
freedom of the electromagnetic fields. We choose the usual gauge  
fixing condition for the Chern-Simons-action, i.e. $A_0 = 0$ for the  
action (3). Thereafter, the Chern-Simons-term in the action reduces  
to the following one

\begin{equation}
- \frac{\sigma_H}{4\pi}\int dt \int_{\Sigma} \epsilon^{mn}  
\dot{A}_m A_n \;\;,
\end{equation}
\label{qact}

with ${\displaystyle\dot{A}_m = \frac{dA_m}{dt}}$ , while the  
$A_0$-term disappears from the Schroedinger-action.

The equations of motion for $A_m$-potentials which result from this  
action beyond the Schroedinger equation $ (i\hbar\partial_t -  
\frac{1}{2\mu} (-i\hbar\partial_m - e A_m )^2 )\psi = 0$ are

\begin{equation}
\frac{ie\hbar}{2\mu} {[ (\partial_m \psi^*)\psi - \psi^*  
(\partial_m\psi)] - \frac{e^2}{\mu} A_m\psi^*\psi} = -  
\sigma_H\epsilon^{mn}\dot{A}_n
\end{equation}
\label{eqmot}

According to the conventional theory \cite{one} the left hand side  
in (5) is the usual current density of the electrons in a magnetic  
field $j_m$ without spin effect which are those also involved in the  
Ohm-equations describing the IQHE. Thus, we obtain the  
Ohm-equations (1) as the equations of motion from the action (3) but  
yet with non-specified $\sigma_H$.

Moreover, the motion or the phase space of system which is  
represented by the action (3) is constrained by the following  
constraint

\begin{equation}
     -\sigma_H \epsilon^{mn}\partial_m A_n = e\psi^*\psi \,\,,
\end{equation}
\label{contraint}

with $ e\psi^*\psi:= j_0$, where $\psi^*\psi$ is the local density  
of carriers.

If we integrate the relation (5) over the sample surface and  
consider $B :=\epsilon_{nm} \partial_m A_n$ as a constant field  
strength, then we obtain the well known relation between the  
Hall-conductivity and the magnetic field, namely

\begin{equation}
{\displaystyle\sigma_H = \frac{ne}{B}} \;\;\;,
\end{equation}
\label{contraint2}

which is the global description of the constraint of our  
microscopic model, where $n = (a)^{-1}\int da (\psi^*\psi)$ is the  
global density of charge carriers which is introduced above and  
$(a)$ is the area of sample.

Accordingly, the continuity equation $\partial_{\alpha}j^{\alpha} =  
0$ is fullfiled in view of the metric $\eta_{\alpha\beta} = ( +1,  
-1, -1 )$ for $M$.

It is worth mentioning that according to the constraint (5) the  
$A_m$-potentials are, generally, usual potentials with non-vanishing  
field strength. However, if the global density of electrons $n$  
becomes negligable which occurres in the typical IQHE-cases [3],  
then $A_m$-potentials become almost pure gauge potentials in view of  
the constraint (5). According to our model, it is this circumstance  
in the typical IQHE-conditions [3] which leads, in view of the pure  
gauge potentials, to the preference of the edge currents occurring  
in the IQHE [3]. We discuss this concept in detail after the  
quantization procedure.

\medskip
Now we show that under the mentioned typical IQH-conditions \cite{vkzwei}
the classical Hall-system becomes a QH-system which must be  
described by a quantized theory. However, in wiew of the fact that  
we consider the IQHE, i.e. a system of carriers with proper low  
mobility and lower density \cite{inteig} \cite{vkzwei} we are  
concerned, as it is well known, with a {\it non-interacting} system  
of carriers \cite{inteig}\cite{Buk}.
In this case which is in view of the non-interacting particles  
quantum mechanically reduciable to a one particle state, the  
schroedinger wave functions of the model should be considered in the  
first quantization. Whereas, an {\it interacting} system of  
carriers with higher mobility and higher magnetic field, i. e. under  
suitable conditions [1] \cite{Buk}, will results in a FQHE-system,  
where the wave functions have to be considered in the second  
quantization \cite{jak}. Thereafter, the wave functions become  
multivalued which results in the fractional quantization of the  
Hall-conductivity \cite{N}.

The main empirical differences between the IQHE-system [3]  
considered in our model and a FQHE-system should be related with the  
lower mobility and strength of the exterior magnetic fields in the  
IQHE case with respect to thevery higher values in the FQHE case [1]  
\cite{Buk}. Of course the difference in magnetic fields is also  
related reciprocally to the density difference between the two cases  
[1]. It is this different experimental setting of IQHE [1], [3]  
with relatively lower mobility and lower magnetic field or higher  
2-D carrier density with respect to the FQHE-case which should  
result in a {\it non-interacting} system of carriers with  
electromagnetic coupling; which further results in the integer  
quantization of the Hall-conductivity according to the single  
valuedness of the Chern-Simons-wave function of the electromagnetic  
potentials in the first quantization.

\bigskip
To quantize the electromagnetic potential $A_m$ one shall use the  
canonical quantization of the Chern-Simons-action (3), from where  
the postulat of quantization can be red off directly \cite{witten  
and}

\begin{equation}
\left[A_m(x_l,t) \,, \, A_n(y_l,t) \right] =  
\frac{4{\pi}i\hbar}{\sigma_H}\,\,\epsilon_{mn}\delta^2( X - Y )    
\;\; ; \: X,Y \in \Sigma\;\;\;\;,
\end{equation}
\label{comutat}

It is convinient to introduce for practical use a Schroedinger  
representation like (7).

We introduce here a new typ of Schroedinger representation of the  
Chern-Simons-theory \cite{mein} which remains very close to the  
canonical quantization.

To do so, one shall use the SO(2)-symmetry of the  
Chern-Simons-action (3) to achive the Schroedinger-representation of  
the quantized Chern-Simons-action by the eigenfunctions of the  
quantum orbital angular momentum operator of the system. This  
operator which is related to the so called holonomy operator in the  
usual quantization of the Chern-Simons-theory \cite{witten and} is  
the generator of the SO(2)-transformation of the action, thus its  
conservation is a consequence of the symmetry of the action (3).

\bigskip
Considering the following canonical transformation in the phase  
space of the Chern-Simons-action (4), i. e. $A_1 = R \cos \phi , A_2  
= R \sin\phi$, the quantum orbital angular operator of the system  
described by the action (3) takes the form $\hat{L} =  
-i\hbar\partial_{\phi}$ \cite{one}. Therafter, we obtain according  
to \cite{one} for the state functions of system (3) the  
eigenfunctions of $\hat{L}$:

\begin{equation}
\Psi (A) = F(R)\, e^{{\;\displaystyle\frac{i}{\hbar} \sigma_H  
l\phi}} \;\;,
\end{equation}
\label{psi}

where F(R) is an arbitrary function of R and the $l = R^2$ is the  
value of the angular momentum of the system which is a constant of  
motion according to the symmetry of the system.

Accordingly, $\sigma_H$ must be quantized in view of the  
single-valuedness
of $\Psi (A)$. Normalizing the constant $R^2 = 1$ and restricting  
us to the positive values the $\sigma_H$ becomes

\begin{equation}
 \sigma_H = 0, 1, 2, ..., N \,; N\in\mathbf Z_+
\end{equation}
\label{sigma}

The integer valuedness of the normalization parameter $k$ of $k  
S_{(C-S)}$ which is in our case $\sigma_H$, as a result of the {\it  
first} quantization of $\sigma_H S_{(C-S)}$ in $\Psi(A)$ on multiply  
valued $\Sigma$ with $(g\geq 1)$ is a well known fact which is  
discussed in various papers very intensively. For a detailed  
discussion of this point for $U(1)$-case see the last contribution  
in Ref. \cite{witten and}.

Thereafter, inserting (10) into (4) one obtains the Ohm-equations  
(1) of IQHE with quantized Hall-conductivity as the equations of  
motion from the
Schroedinger-(quantum) Chern-Simons-action (3) with (8)-(10).

Thus, as it is discussed above, the typical experimental setting of  
the IQHE under the typical IQH-conditions \cite{vkzwei} results in  
the {\it non-interacting } system of carriers with electromagnetic  
coupling which is described by a single valued wave function. It is  
this chain of correlated properties which implies the integer  
quantization of the $\sigma_H$.

\medskip
Recall also that the normalization parameter of the $\Psi_{C-S}$  
becomes allways quantized as integers in view of the single  
valuedness of $\Psi_{C-S}$ in its first quantization no matter what  
kind of quantization is performed \cite{witten and} \cite{jak}.

Of course we are aware about the question of the massive photons in  
a quantized 2+1-dimensional theory
\cite{deser}. The question has however in our model a favorable answere. 

First of all, the mass parameter usually discussed in the  
2+1-dimensional case arises from a comparison between the two  
available action terms for the electromagnetic potential in this  
dimension: The Maxwell-term $L\int F_{\alpha\beta} F^{\alpha\beta}$  
and the Chern-Simons-term
$ K\int \epsilon^{\alpha\beta\gamma} A_{\alpha}\partial_{\beta}  
A_{\gamma}$ \cite{deser}, where K must be a dimensionless parameter.  
The mass parameter is then given by $M \sim (K L^{-1})$.
It is the dimensional structure of the Maxwell-term in the  
2+1-dimensions which forces to be coupled with a parameter of the  
order $L\sim M^{-1}$. The Chern-Simons-term itself does not need any  
dimensional coupling, hence the $\sigma_H$ is in our model also  
dimensionless. Therefore, since we do not use the Maxwell-term in  
our action (2), so we have no such mass term and in view of its  
dimensionless character the $\sigma_H$ alone can not play the role  
of mass parameter.

\bigskip
We mentioned already that most of experimental data related to the  
IQHE can be understood in a satisfactory manner, if the currents  
involved in IQHE are considered as the adge currents \cite{vkzwei}.  
On the ohter hand, according to  the Ohm-equations $j_m =  
\epsilon_{nm}\sigma_H \partial_t A_n$ the currents
are restricted to those regions where the $A_m$-potentials are  
allowed to exist. Thus, the question of the edge currents is related  
with the questions of the regions where the $A_m$-potentials are  
allowed to exist.

For a theoretical conception of the edge currents in this model we  
use the appearence of pure gauge potentials, i e. those with  
vanishing fiels strength, in our model. This circumstance take  
place, if we adopt in the model, i. e. in the constraint (5) the  
experimental occasions which are typical for the IQHE-cases [3].

Fortunately, in this model the only allowed $A_m$-potentials are  
those existing on the boundary of region $\Sigma$, if we consider a  
negligable density of electrons with respect to the megnitude of  
very high magnetic fields (in accordance with the typical  
QHE-arrangements) [3]. This becomes obvious if one recall that the  
$A_m$-potentials are constrained to be pure gauge potentials in  
those regions, where according to the constraint (5) the density of  
electrons is negligable. This is the case in typical  
IQHE-experiments with small device currents \cite{vkzwei} or on the  
boundaries of samples and also it is the case specially within the  
impurity regions or the Corbino-disc, where the dencity of electrons  
can be considered to be almost zero. Recall also that conversely at  
large transport currents the IQHE breaks down \cite{vkzwei}.

Thus, in view of these circumstances in typical IQHE-cases we may  
replace the constraint (5) by the following one

\begin{equation}
           \epsilon^{mn}\partial_m A_n \approx 0 \;\;\;,
\end{equation}
\label{cosnt3}

Thereafter, the $A_m$-potentials become almost pure gauge  
potentials, i.e. $A_m \approx ig^{-1} \partial_m g$, where g is an  
element of the U(1)-gauge group.

On the other hand, the constraint tensor $\epsilon_{mn}\partial_m  
A_n$ generates a gauge transformation $A_m^\prime = A_m +\partial_m  
\lambda$ in the phase space of the $A_m$-potentials \cite{witten  
and}.

Therefore, according to the constraint (11) one must identify  
$A_m^\prime = A_m$ everywhere in the phase space.

Furthermore, if as in our case the $\Sigma$ possess a boundary we  
must  choose boundary conditions for $A_m$ and $\lambda$ on the  
boundary. We choose free boundary conditions for $A_m$ but $\lambda  
= 0$ on the boundary. A reason for this choise is that the  
Chern-Simons-action is not invariant under gauge transformations  
that do not vanish on the boundary \cite{witten and}.

Accordingly, we must identify $A_m^\prime = A_m$ for any $\lambda$  
which vanishes on the boundary $\partial\Sigma$. The only pure $A_m$  
gauge potentials which obey this additional condition are those  
restricted to be defined only on the boundary \cite{witten and}. In  
other words, the only $A_m$-potentials obeying both restrictions  
caused by the constraint (11) are those restricted to exists on the  
boundary region of $\Sigma$. Thereafter, the currents $j_m$ should  
be considered also to be restricted to the boundary region, i. e. to  
the so called edge currents. This conception can be considered as a  
theoretical background for the preference of the edge currents in  
the typical IQHE experiments \cite{vkzwei}.

\bigskip
In conclusion we may mention however, that the IQHE is also  
observed in the absence of edge currents \cite{neu} and moreover  
that at large transport currents the IQHE breaks down \cite{vkzwei}.  
To understand these questions within our model one may recall the  
following characteristics of the model.

As it is obvious from the model the quantization of $\sigma_H$  
depends only on the canonical quantization of the  
Chern-Simons-action for usual gauge potentials; Whereas, the  
occurrance of edge currents is a result of the boundary conditons  
adopted for $\lambda$ in the case of pure gauge potentials.

It results from the Ohm-equations with quantized $\sigma_H$, as the  
equations of motion in our model, that both components of the $j_m$  
current density are proportional to the electric fields
by the quantized $\sigma_H$. Thus, although one of the electric  
field components should be turned off in view of some experimental  
arrangement \cite{neu}, nevertheless this circumstance have no  
influence on the quantization of $\sigma_H$ in our model.

Moreover, vanishing of an electric field component in a region does  
not mean that the related component of the electrmagnetic potential  
in the same region vanishes also. In our model considering the $A_0  
= 0$ gauge fixing it means that the mentioned component of the  
potential is konstant in time which is in accordance with its pure  
gauge potential character. Thus, as long as the action of the  
$A_m$-potentials ( at least its kinetic term) is described by the  
Chern-Simons-action, the quantization of the $\sigma_H$ according to  
our model is straightforward also for the case, where one of the  
components of the electric field vanishes.

\bigskip
On the other hand, to understand the break down of the IQHE  
occurring at large transport currents one shall recall that the  
topological character of the IQHE in our model is closely related  
with the topological invariance of the Chern-Simons-term for self  
interaction of the pure electromagnetic gauge potentials. However,  
if the density of electrons is not negligable which should be the  
case at large transport currents, then the $A_m$-potentials are not  
more forced to be pure gauge potentials in view of the constraint  
(5) or (11). Thus, they become usual potentials with non-vanishing  
field strength. Therefore, not only that the Chern-simons-term alone  
is no more adequate to describe the dynamics of such potentials but  
also the typical topological character of the line integrals of  
pure U(1)-gauge potentials which is suitable for the topological  
character of IQHE is lost in this case. Accordingly, in this case in  
view of the absence of the pure gauge potentials the typical edge  
currents of IQHE become also absent. Furthermore, in this case  
either the Chern-Simons-term must be absent in the kinetical part of  
the action or there must be additional  Maxwell-term to describe  
the dynamics of the usual electromagnetic potentials. However, then  
the typical quantization of the pure Chern-Simons-term which is  
showed to result in the quantized $\sigma_H$ is no more possible.

In this manner, the model can be helpful to understand both of the  
mentioned questions.

The further questions of vanishing longitudinal conductivity in the  
IQHE and of the fractional FQHE will be investigated elsewhere  
\cite{under}.

After preparing this paper we are informed that similar ideas can  
be found in recent works by J. Froehlich and coauthors \cite{FRO},  
where they use different path-integral and also lattice theoretical  
methodes.

\newpage
Footnotes and references


\begin{thebibliography}{100}
\bibitem{erklar}
Beyond the geometrical significance of boundary to define  
potentials according to the differential geometry on real samples  
(regions) with boundary it becomes also important in view of the  
boundary conditions for field theories. See furthermore  
\cite{schwarz}
\bibitem{diff}
This means that $d\sigma_H = 0$. Recall that, for example if we  
work with a constant n but a variable B-field which is the case in  
some of the QHE-experiments then the Hall conductivity $\sigma_H =  
{\displaystyle\frac{en}{B}}$ is given by  
${\displaystyle\frac{en}{dA}}$ with $B$ as the component of $dA =  
(\partial_1 A_2 - \partial_2 A_1) dx^1 \wedge dx^2$, which is  
obviously locally constant in view of $d^2 = 0$. For further  
arguments in favour of the local constancy of $\sigma_H$ see Ref.  
[1f].
Furthermore, recall also that both local and global charge carier  
dencities in two dimensions and the B-field are of dimension  
$L^{-2}$.Thus, the $\sigma_H = {\displaystyle\frac{en}{B}}$ becomes  
dimensionles.
\bibitem{jak}
See for other methodes of the first quantizations of $\Psi(A)$ Ref.  
\cite {witten and}. For the second quantization of $\Psi(A)$ which  
results in the multivalued wave functions see the models quoted in  
Ref. \cite{N}. The fractional quantization of the normalization  
parameter, which is belived to be related with fractional Hall  
conductivity in FQHE, should be a result of multivaluedness of  
$\Psi(A)$in its second quantization which is related with the  
interacting electrons (see the models cited in Ref. \cite{N} and see  
also Ref. \cite{Buk}).
\end{thebibliography}
\end{document}